\documentclass[12pt]{article}

\usepackage{amssymb}
\begin{document}
%
\begin{flushright}
{December, 2010}
\end{flushright}
\vspace{0.1cm}
\begin{center}
{\Large \bf
A quantum isomonodromy equation and its application to ${\cal N}=2$ $SU(N)$ gauge theories
}
\vskip0.5cm
{\large Yasuhiko Yamada}
\vskip0.5cm
{\it 
Department of Mathematics, Faculty of Science \\
Kobe University, Hyogo, 657-8501, Japan}
\end{center}
\vskip0.5cm

\begin{abstract}
We give an explicit differential equation which is expected to determine the instanton partition function
in the presence of the full surface operator in ${\cal N}=2$ $SU(N)$ gauge theory.
The differential equation arises as a quantization of a certain Hamiltonian system of isomonodromy type
discovered by Fuji, Suzuki and Tsuda.
\end{abstract}

{\footnotesize PACS: 11.25.HF, 11.15.-q.

}


\section{Introduction}

In \cite{Alday:2010vg}, Alday and Tachikawa formulated a combinatorial formula\footnote{We will recapitulate 
it in section 2, eq.(\ref{eq:Zinst}).} for the instanton partition function 
$Z_{\rm inst}$ in the presence of the full surface operator in ${\cal N}=2$ $SU(N)$ gauge theory, based on a 
result \cite{Feigin} about the affine Laumon space. 
Furthermore, they observed an interesting relation between $Z_{\rm inst}$ for $SU(2)$ theory and KZ equation 
with affine $SL(2)$ symmetry as an extension of the AGT relation \cite{Alday:2009aq}.
A similar relation to Virasoro CFT including the irregular singularities was then examined in \cite{MT}\cite{AFKMY}.
For $SU(N)$ case, the relation to affine $SL(N)$ conformal blocks was studied in \cite{Kozcaz:2010yp}.

In this note, we will study the differential equation satisfied by $Z_{\rm inst}$ from a 
point of view slightly different from KZ equations or $W_N$-algebras.
Our basic strategy is to use the isomonodromy equations.
It is known that some aspects of AGT relation have a natural interpretation \cite{AFKMY}\cite{tesch}\cite{Tai} through the
isomonodromy systems (or Painlev\'e equations for $SL(2)$ cases). That is,
the $2d$ CFT's (=quantum isomonodromy systems) can be viewed as
non-autonomous and quantum deformation of the Hitchin systems \cite{BD}\cite{BT} (=Seiberg-Witten theory)
arising through the $\Omega$-deformation \cite{NW}.

At the classical level, the relation between the
Seiberg-Witten theory and isomonodromy equation is directly recognized by looking at
the curves. For instance, 
the $SU(2)$ $N_f=4$ Seiberg-Witten curve in Gaiotto form \cite{Gaiotto:2009we}
\begin{equation}
x^2
=(\frac{\mu_1}{z}+\frac{\mu_2}{z-1}+\frac{\mu_3}{z-t})^2-\frac{\kappa z+u}{z(z-1)(z-t)},
\quad \kappa=\mu_0^2-(\sum_{i=1}^3 \mu_i)^2,
\end{equation}
coincide with the Hamiltonian of the sixth Painlev\'e equation
\begin{equation}
\begin{array}l
H_{\rm VI}=q(q-1)(q-t)p^2-2 \{\mu_1(q-1)(q-1)+\mu_2q(q-t)+ \mu_3q(q-1)\}p-\kappa q,
\end{array}
\end{equation}
by the following variable change
\begin{equation}
\begin{array}l
\displaystyle p=x+\frac{\mu_1}{z}+\frac{\mu_2}{z-1}+\frac{\mu_3}{z-t}, \quad q=z, \quad H_{\rm VI}=u.
\end{array}
\end{equation}
Similar relations for degenerate cases were considered in \cite{KMNOY}.

We want to generalize this kind of correspondence to $SU(N)$ $N_f=2N$ cases at the quantum level. The first problem is to look for
the suitable isomonodromy system with higher rank symmetries. Fortunately, a nice candidate appeared in recent work \cite{Tsu1}\cite{Suz}.
The system, which we call
Fuji-Suzuki-Tsuda (FST) equation\footnote{This equation (called $P_{\rm VI}$-chain in \cite{Tsu1})
was first considered by Tsuda in 2008, as a similarity reduction of his 'UC-hierarchy' (certain generalization of KP hierarchy).
Independently, in the context of the Drinfeld-Sokolov hierarchy, it was obtained by Fuji-Suzuki \cite{FS} in case of $N=3$ and generalized in \cite{Suz}. 
Though our description in section 3 will follow the notation of \cite{Suz},
the isomonodromic picture considered here is naturally understood from the UC-hierarchy.
Tsuda's construction contains also the Garnier type extension with spectral type
$(N-1,1),\cdots,(N-1,1),(1^N),(1^N)$ (see \cite{Tsu2}).}, can be described as an isomonodromy deformation of the $N \times N$
Fuchsian connection on ${\mathbb P}^1$ with regular singularity at $z=0,1,t,\infty$
\begin{equation}\label{eq:conn}
D={\partial}_z-{\cal A}, \quad {\cal A}=\frac{A_0}{z}+\frac{A_1}{z-1}+\frac{A_t}{z-t},
\end{equation}
with the following spectral type (= eigenvalue multiplicity of the residue matrices):
$A_0$ and $A_{\infty}=-A_0-A_1-A_t$ are of type $(1^N)$ while
$A_1$ and $A_t$ are of type $(N-1,1)$ \footnote{Then, without loss of generality, one can assume that the eigenvalues
of $A_1$ are $(0,\cdots,0,c)$ and similar for $A_t$.}.
We will recall the explicit form of the isomonodromy equation in section 3.
Here, we look at the classical spectral curve in order to see its relation
to the gauge theory.\footnote{
Another indication comes from the special solutions. In both theories,
generalized hypergeometric series ${}_NF_{N-1}$ appears \cite{Suz-sol}\cite{Tsu-sol}, \cite{MM}\cite{ScWy}\cite{Awata:2009ur}.} The curve for eq.(\ref{eq:conn}) is given by 
\begin{equation}
\begin{array}l
\det(v-z {\cal A})\propto\det\Big((z-1)(z-t)(v-A_0)-z(z-t)A_1-z(z-1)A_t\Big)\\
\phantom{\det(v-z {\cal A})}=\{(z-1)(z-t)\}^{N-1}f(z,v)=0,
\end{array}
\end{equation}
where $v=z\partial_z$ is considered as a commuting variable.
The factorization in the second line is due to the fact that $A_1$, $A_t$ are of rank one.
Hence $f(z,v)$ is of bi-degree $(2,N)$ in variables $(z,v)$ and has the form
\begin{equation}
f(z,v)=t\prod_{i=1}^N (v-m_i)+z\Big\{\sum_{i=0}^{N-1} u_i v^i-(1+t)v^N\Big\}+z^2\prod_{i=1}^N (v-{\tilde m}_i),
\end{equation}
where $A_0 \sim {\rm diag}(m_1,\cdots, m_N)$, $A_{\infty} \sim -{\rm diag}({\tilde m}_1,\cdots, {\tilde m}_N)$.
This is the desired form as the Seiberg-Witten curve for $SU(N)$ with $N_f=2N$.

In the next section, we formulate our main conjecture that a differential equation
(\ref{eq:main-de}) determines the instanton partition function $Z_{\rm inst}$.
The isomonodromic origin of our equation (\ref{eq:main-de}) is discussed in section 3.

\section{Main conjecture}

The instanton partition function $Z_{\rm inst}=Z_{\rm inst}(y;{a},{m},{\tilde {m}})$
in the presence of the full surface operator in $SU(N)$ $N_f=2N$ 
superconformal gauge theory
is a function of $N$-variables ${y}=(y_1,\cdots,y_N)$ depending on $3N-1$ parameters
${m}=(m_1,\cdots,m_N)$, ${\tilde {m}}=({\tilde m}_1,\cdots,{\tilde m}_N)$ and ${a}=(a_1,\cdots,a_N)$,
$\displaystyle \sum_{i=1}^N a_i=0$.

Let us recall the combinatorial formula 
for $Z_{\rm inst}$ following \cite{Alday:2010vg}\cite{Kozcaz:2010yp}
\begin{equation}\label{eq:Zinst}
Z_{\rm inst}
=\sum_{\lambda} Z(\lambda)\prod_{i=1}^N y_i^{k_i(\lambda)}.
\end{equation}
Here the sum is taken over all $N$-tuples $\lambda=(\lambda^1, \cdots, \lambda^N)$
of partitions $\lambda^i=(\lambda^i_1\geq \lambda^i_2\geq \lambda^i_3\geq \cdots \geq \lambda^i_{\ell_i}> 0)$. 
The indices of $\lambda^i=(\lambda^i_j)$ will be extended to $\Bbb Z$ by $\lambda^i=\lambda^{i+N}$ and
$\lambda^i_j=0$ ($j\leq 0$ or $j>\ell_i$). We put 
$\displaystyle |\lambda|=\sum_{i=1}^N |\lambda^i|=\sum_{i=1}^N \sum_{j\geq 1} \lambda^i_j$.
The exponents $k_i(\lambda)$ are given by
\begin{equation}
k_i(\lambda)=\sum_{j\geq 1}\lambda^{i-j+1}_j.
\end{equation}
The coefficients $Z(\lambda)$ are defined as
\begin{equation}
Z(\lambda)=\frac{n_f(a,\lambda,m)n_{\tilde f}(a,\lambda,\tilde{m})}{n_v(a,\lambda)},
\end{equation}
where
\begin{equation}\label{eq:subs}
\begin{array}l
n_f(a,\lambda,m)=n_{\rm bif}(m,a,(\phi)^N,\lambda,0),\\
n_{\tilde f}(a,\lambda,{\tilde m})=n_{\rm bif}(a,{\tilde m},\lambda,(\phi)^N,0),\\
n_v(a,\lambda)=n_{\rm bif}(a,a,\lambda,\lambda,0),\\
\displaystyle
n_{\rm bif}(a,b,\lambda,\mu,x)=\prod_{t=1}^{|\lambda|+|\mu|}(w_t-x),
\end{array}
\end{equation}
and finally, the weights $w_t=w_t(a,b,\lambda,\mu)$\footnote{$a, b \in {\Bbb C}^N$, and $\lambda, \mu$ are
$N$-tuples of partitions. The constraints $\sum_{i=1}^N a_i=0$, $\sum_{i=1}^N b_i=0$
will be considered after the substitutions (\ref{eq:subs}).}
are determined by the formula \cite{Feigin}

\begin{equation}\label{eq:weight}
\begin{array}{l}
\displaystyle
\chi(a,b,\lambda, \mu)=\sum_{t=1}^{|\lambda|+|\mu|}e^{w_t}\\
=\displaystyle
\sum_{k=1}^N \sum_{l'\geq 1} e^{a_{k}-b_{k-l'}+\epsilon_2(\lfloor \frac{{l'}-k}{N} \rfloor-\lfloor \frac{-k}{N} \rfloor)}
\sum_{s=1}^{\mu^{k-l'}_{l'}} e^{\epsilon_1 s}\\
-\displaystyle
\sum_{k=1}^N \sum_{l\geq 1}\sum_{l'\geq 1}
e^{a_{k-l+1}-b_{k-l'}+\epsilon_2(\lfloor\frac{{l'}-k}{N} \rfloor-\lfloor\frac{l-k-1}{N} \rfloor)}
(e^{\epsilon_1 \mu^{k-l'}_{l'}}-1)
\sum_{s=1}^{\lambda^{k-l+1}_l}e^{\epsilon_1(s-\lambda^{k-l+1}_l)}\\
+\displaystyle
\sum_{k=1}^N \sum_{l\geq 1}\sum_{l'\geq 1}
e^{a_{k-l+1}-b_{k-l'+1}+\epsilon_2(\lfloor\frac{{l'}-k-1}{N} \rfloor-\lfloor\frac{l-k-1}{N} \rfloor)}
(e^{\epsilon_1 \mu^{k-l'+1}_{l'}}-1)
\sum_{s=1}^{\lambda^{k-l+1}_l}e^{\epsilon_1(s-\lambda^{k-l+1}_l)}\\
+\displaystyle
\sum_{k=1}^N \sum_{l\geq 1}
e^{a_{k-l+1}-b_{k}+\epsilon_2(\lfloor\frac{-k}{N} \rfloor-\lfloor\frac{l-k-1}{N} \rfloor)}
\sum_{s=1}^{\lambda^{k-l+1}_l}e^{\epsilon_1(s-\lambda^{k-l+1}_l)},
\end{array}
\end{equation}
where $\lfloor x \rfloor$ is the largest integer such that $\lfloor x \rfloor\leq x$.
The formula (\ref{eq:weight}) is applicable to periodic parameters $a_{i+N}=a_i$, $b_{i+N}=b_i$.
However, another convention is possible, where all terms $\epsilon_2(\lfloor \cdots \rfloor-\lfloor \cdots \rfloor)$
in the exponentials in eq.(\ref{eq:weight}) are taken away in exchange for assuming quasi-periodicity 
$a_{i+N}=a_i+\epsilon_2$, $b_{i+N}=b_i+\epsilon_2$.
In what follows, we will adopt the second option.

\medskip
As compared with the above complicated formulae for $Z_{\rm inst}$, the differential equation 
we propose is rather simple and defined as
\begin{equation}\label{eq:main-de}
\begin{array}l
\displaystyle
{\cal D}{\cal Z}(y)=\Big\{(\Delta_N+\sum_{i=1}^N u_i \vartheta_i)+
(\prod_{j=1}^{N} y_j)(\Delta_N+\sum_{i=1}^N v_i \vartheta_i+\sum_{i=1}^N r_is_i)\\
\displaystyle
+\sum_{i=1}^N (y_i+y_iy_{i+1}+\cdots+\prod_{j=0}^{N-2}y_{i+j})
(\vartheta_{i-1}-\vartheta_i+r_i)(\vartheta_{i-1}-\vartheta_i+s_i)\Big\}{\cal Z}=0,
\end{array}
\end{equation}
where $y_{i+N}=y_i$, $\displaystyle \vartheta_i=y_i\frac{\partial}{\partial y_i}$, 
$\displaystyle \Delta_N=\frac{1}{2} \sum_{i=1}^N (\vartheta_i-\vartheta_{i+1})^2$.
The parameters $u_i, v_i, r_i, s_i$ $(1\leq i\leq N)$ will be set as
\begin{equation}\label{eq:para-set}
\begin{array}l
\displaystyle
u_i=\frac{a_{i+1}-a_i}{\epsilon_1}, \quad
v_i=\frac{a_{i+1}-a_i+m_{i+1}-m_{i+2}+{\tilde m}_{i}-{\tilde m}_{i+1}}{\epsilon_1}, \\
\displaystyle
r_i=\frac{a_i-m_{i+1}-\epsilon_1}{\epsilon_1}, \quad
s_i=\frac{a_i-{\tilde m}_i}{\epsilon_1}, 
\end{array}
\end{equation}
where $x_{i+N}=x_i+\epsilon_2$ for $x=a, m, {\tilde m}$ (while $u_i,v_i,r_i,s_i$ are periodic).

The main claim in this note is the following

\noindent
{\bf Conjecture.}
{\it The instanton partition function $Z_{\rm inst}$ is characterized as the unique formal power
series solution of the form ${\cal Z}=1+{\cal O}(y)$ for the differential equation (\ref{eq:main-de}) with 
parameters (\ref{eq:para-set}).}

\medskip
We have checked this conjecture for $N\leq 5$ up to total degree $5$ in $y$-variables (in some cases
by specializing the parameters to numerical values).

Under a degeneration limit $y_i \rightarrow \varepsilon^2 y_i$, $m_i ({\tilde m}_i) \rightarrow \varepsilon^{-1} \Lambda$, ($\varepsilon \rightarrow 0$), the differential equation (\ref{eq:main-de}) 
reduces to the Toda equation 
\begin{equation}
\Big(\Delta_N+\sum_{i=1}^N u_i \vartheta_i+\frac{\Lambda^2}{\epsilon_1^2}\sum_{i=1}^N y_i\Big){\cal Z}=0,
\end{equation}
whose relation to ${\cal N}=2$ $SU(N)$ pure gauge theory
has already been established by Braverman-Etingof \cite{Braverman:2004cr} (see also \cite{FFFR}\cite{Negut}\cite{BFRF}).

\section{Origin of the differential equation}

In this section, we explain an isomonodromic origin of our differential equation (\ref{eq:main-de}).
As already mentioned in the introduction, the equation (\ref{eq:main-de}) is a quantization of 
the Fuji-Suzuki-Tsuda (FST) equation \cite{FS}\cite{Tsu1}\cite{Suz}.
The FST equation can be written as a Hamiltonian system for $2(N-1)$ variables $q=(q_1,\cdots,q_{N-1})$ and 
$p=(p_1 \cdots, p_{N-1})$ 
\begin{equation}
t(t-1)\frac{dq_i}{dt}=\frac{\partial H}{\partial p_i}, \quad
t(t-1)\frac{dp_i}{dt}=-\frac{\partial H}{\partial q_i},
\end{equation}
with parameters
$\eta$ and $\alpha=(\alpha_0, \cdots, \alpha_{2N-1})$, $\displaystyle \sum_{j=0}^{2 N-1} \alpha_j=1$.
The Hamiltonian $H=H(q,p,t;\eta,\alpha)$ is given by\footnote{This is a kind of coupled system of 
the Painlev\'e VI. Another equation
of such type first found by Sasano \cite{Sasano} will also be important for some superconformal gauge theories.} 
\begin{equation}
\begin{array}l
\displaystyle
H=\sum_{i=1}^{N-1} H_{\rm VI}(q_i,p_i;a_i,b_i,c_i,d_i)\\
\displaystyle
\qquad +\sum_{1\leq i<j\leq N-1}(q_i-1)(q_j-t)\{(q_ip_i+\alpha_{2 i-1})p_j+p_i(q_jp_j+\alpha_{2 j-1})\},
\end{array}
\end{equation}
where $H_{\rm VI}$ is the Hamiltonian of sixth Painlev\'e equation
\begin{equation}
H_{\rm VI}(q,p;a,b,c,d)=q(q-1)(q-t)p^2 -\{aq(q-1)+bq(q-t)+c(q-1)(q-t)\}p+dq,
\end{equation}
and
$\displaystyle a_i=\sum_{j=i}^{N-1}\alpha_{2j}$, 
$\displaystyle b_i=\sum_{j=0}^{i-1}\alpha_{2j}$, 
$\displaystyle c_i=\sum_{j=0}^{N-1}\alpha_{2j+1}-\alpha_{2i-1}-\eta$,
$d_i=\eta \alpha_{2i-1}$.

A quantization of the system is given by the Schr\"odinger equation\footnote{Here we will not consider the problem of operator ordering seriously
since the ambiguities can be absorbed by shifts of parameters.}
\begin{equation}\label{eq:Schro}
\Big\{t(t-1)\frac{\partial}{\partial t} -H(q_i,\frac{\partial}{\partial q_i})\Big\}{\tilde \Psi}(q_1,\cdots,q_{N-1},t)=0.
\end{equation}
To connect this equation with $Z_{\rm inst}$, we make a gauge transformation
$
{\tilde \Psi}(q,t)=y_1^{k_1}\cdots y_N^{k_N} \Psi(y),
$
together with the variables change
\begin{equation}
\begin{array}l
\displaystyle
y_1=q_1,\quad y_2=\frac{q_2}{q_1},\quad \cdots,\quad y_{N-1}=\frac{q_{N-1}}{q_{N-2}},\quad
y_N=\frac{t}{q_{N-1}},\\
\displaystyle
q_i\frac{\partial}{\partial q_i}=\vartheta_i-\vartheta_{i+1} \quad (i=1,\cdots,N-1), \quad
t\frac{\partial}{\partial t}=\vartheta_N,
\end{array}
\end{equation}
($\vartheta_i=y_i \frac{\partial}{\partial y_i}$).
Then the eq.(\ref{eq:Schro}) takes the form 
(in the following, we will concentrate on $N=3$ case)
\begin{equation}\label{eq:N3prime}
\begin{array}l
{\cal D}' \Psi(y)=\{\Delta_3+u_1\vartheta_1+u_2\vartheta_2+u_3\vartheta_3+M_0\\
+y_1(\vartheta_{31}+s_1)(\vartheta_{12}+r'_1)
+y_2(\vartheta_{12}+s_2)(\vartheta_{23}+r'_2)
+y_3(\vartheta_{23}+s_3)(\vartheta_{31}+r'_3)\\
+y_2y_3(\vartheta_{12}+s_2)(\vartheta_{31}+r'_3)
+y_3y_1(\vartheta_{23}+s_3)(\vartheta_{12}+r'_1)
+y_1y_2(\vartheta_{31}+s_1)(\vartheta_{23}+r'_2)\\
+y_1y_2y_3(\Delta_3+v_1\vartheta_1+v_2\vartheta_2+v_3\vartheta_3+M_1)\}\Psi(y)=0,
\end{array}
\end{equation}
where $\vartheta_{ij}=\vartheta_i-\vartheta_j$, and $r'_i,s_i,u_i,v_i,M_0,M_1$ are
some constants depending only on the parameters $\eta, \alpha$ and $k_1,\cdots,k_N$
(their precise expressions are not necessary).
We can and will choose $k_N$ so that $M_0=0$.
Then, the equation (\ref{eq:N3prime}) have a unique formal series solution of the form
\begin{equation}\label{eq:psi-c}
\Psi(y)=\sum_{i,j,k=0}^{\infty} c_{ijk}y_1^iy_2^jy_3^k. \quad (c_{000}=1)
\end{equation}
The coefficients $c_{ij0}$ are written in terms of the very-well-poised, balanced hypergeometric series
\begin{equation}
F(a_0;a_1,\cdots,a_5)=\sum_{k=0}^{\infty}\frac{(a_0+2k)}{a_0}\prod_{i=0}^5 \frac{(a_i)_k}{(a_0+1-a_i)_k},
\end{equation}
as 
\begin{equation}\label{eq:Fsol}
c_{ij0}=\frac{(r'_1-j)_i(r'_2)_j(-s_1)_i(-s_2)_j}{i!j!(u_1+1-j)_i(u_2+1)_j}F(u_1-j;-i,-j,u_1-r'_1+1,u_1-s_2,-u_2-j),
\end{equation}
where $(x)_i=\Gamma(x+i)/\Gamma(x)$ is the Pochhammer symbol (see Appendix A for the proof).
Similarly, $c_{0ij}$ and $c_{j0i}$ are given by cyclic shifts of parameters $x_i\rightarrow x_{i+1 \ ({\rm mod}\  3)}$ ($x=r,s,u$).

On the other hand, corresponding coefficients of the instanton partition function
\begin{equation}
Z_{\rm inst}(y)=\sum_{i,j,k=0}^{\infty} c^{L}_{ijk}y_1^iy_2^jy_3^k,
\end{equation}
are obtained (at least for the first several terms) as
\begin{equation}
\begin{array}l
\displaystyle
c^{L}_{ij0}=\sum_{k=0}^{{\rm min}(i,j)}Z(\{i,k\},\{j-k\},\phi)\\
\displaystyle
=(-1)^{i+j}\frac{(\frac{\epsilon_1-a_1+m_2}{\epsilon_1})_i(\frac{\epsilon_1-a_2+m_3}{\epsilon_1})_j
(\frac{-a_1+{\tilde m}_1}{\epsilon_1})_i(\frac{-a_2+{\tilde m}_2}{\epsilon_1})_j}{i! j! (\frac{\epsilon_1-\epsilon_1 j-a_1+a_2}{\epsilon_1})_i
(\frac{\epsilon_1-a_2+a_3}{\epsilon_1})_j}\\
\quad \times
F(\frac{-\epsilon_1 j-a_1+a_2}{\epsilon_1};-i,-j,\frac{-\epsilon_1 j+a_2-a_3}{\epsilon_1},\frac{\epsilon_1-a_1+m_3}{\epsilon_1},\frac{-a_1+{\tilde m}_2}{\epsilon_1}),
\end{array}
\end{equation}
together with similar formulas for $c^{L}_{0ij}$, $c^{L}_{j0i}$ obtained by shifts $x_1\rightarrow x_2 \rightarrow x_3 \rightarrow x_1+\epsilon_2$
($x=a, m, {\tilde m}$).
Comparing the coefficients $c_{ij0}$ and $c^{L}_{ij0}$ etc., we find that they almost coincide if we put
\begin{equation}
\begin{array}l
\displaystyle
r'_1= \frac{a_2-m_3}{\epsilon_1},\quad
r'_2= \frac{a_3-(m_1+\epsilon_2)}{\epsilon_1},\quad
r'_3= \frac{a_1-m_2}{\epsilon_1},\\
\displaystyle
s_1= \frac{a_1-{\tilde m}_1}{\epsilon_1},\quad
s_2= \frac{a_2-{\tilde m}_2}{\epsilon_1},\quad
s_3= \frac{a_3-{\tilde m}_3}{\epsilon_1},\\
\displaystyle
u_1= \frac{a_2-a_1}{\epsilon_1},\quad
u_2= \frac{a_3-a_2}{\epsilon_1},\quad
u_3= \frac{(a_1+\epsilon_2)-a_3}{\epsilon_1}.
\end{array}
\end{equation}
In fact, under this parameter identification, the ratio of the coefficients $c_{ij0}/c^{L}_{ij0}$ etc. are
simply given by
\begin{equation}\label{eq:norm}
\frac{c_{ijk}}{c^{L}_{ijk}}=(r'_1)_{i-j}(r'_2)_{j-k}(r'_3)_{k-i}.
\end{equation}
Moreover, this relation (\ref{eq:norm}) is satisfied also for $(ijk)=(111), (211), (121), (112)$ and further (as far as we checked), by putting
\begin{equation}
\begin{array}l
M_1=(r'_3-1)s_1+(r'_1-1)s_2+(r'_2-1)s_3, \\
v_1=r'_1-r'_3-s_1+s_2-u_1,\quad
v_2=r'_2-r'_1-s_2+s_3-u_2,\\
v_3=r'_2-r'_1-s_3+s_1-u_3-\frac{\epsilon_2}{\epsilon_1}.
\end{array}
\end{equation}

In order to remove the factor in eq.(\ref{eq:norm}), we make a "gauge transformation" (in momentum space) defined by
\begin{equation}
{\cal D}' \rightarrow {\cal D}= V^{-1} {\cal D}'V, \quad V=(r'_1)_{\vartheta_{12}}(r'_2)_{\vartheta_{23}}(r'_3)_{\vartheta_{31}}.
\end{equation}
Under this transformation, the Euler derivatives remain invariant
$\vartheta_i \rightarrow V^{-1}\vartheta_i V=\vartheta_i$.
On the other hand, by using the relation $\vartheta_iy_j=y_j (\vartheta_i+\delta_{ij})$, the multiplication operators $y_i$ are transformed as
\begin{equation}
\begin{array}l
\displaystyle 
y_1 \rightarrow V^{-1}y_1V
=y_1 \frac{(r'_1)_{\vartheta_{12}}(r'_2)_{\vartheta_{23}}(r'_3)_{\vartheta_{31}}}{(r'_1)_{\vartheta_{12}+1}(r'_2)_{\vartheta_{23}}(r'_3)_{\vartheta_{31}-1}}= y_1 \frac{\vartheta_{31}+r'_3-1}{\vartheta_{12}+r'_1},\\[4mm]
\displaystyle
y_2 \rightarrow y_2 \frac{\vartheta_{12}+r'_1-1}{\vartheta_{23}+r'_2},\quad
y_3 \rightarrow y_3 \frac{\vartheta_{23}+r'_2-1}{\vartheta_{31}+r'_3},
\end{array}
\end{equation}
and hence
\begin{equation}
y_1y_2 \rightarrow y_1 \frac{\vartheta_{31}+r'_3-1}{\vartheta_{12}+r'_1}
y_2 \frac{\vartheta_{12}+r'_1-1}{\vartheta_{23}+r'_2}
=y_1y_2 \frac{\vartheta_{31}+r'_3-1}{\vartheta_{23}+r'_2}.
\end{equation}
Then the differential operator ${\cal D}'$ in eq.(\ref{eq:N3prime}) is transformed into
\begin{equation}
\begin{array}l
{\cal D}=V^{-1} {\cal D'}V=\Delta_3+u_1\vartheta_1+u_2\vartheta_2+u_3\vartheta_3\\
+y_1(\vartheta_{31}+s_1)(\vartheta_{31}+r_1)
+y_2(\vartheta_{12}+s_2)(\vartheta_{12}+r_2)
+y_3(\vartheta_{23}+s_3)(\vartheta_{23}+r_3)\\
+y_2y_3(\vartheta_{12}+s_2)(\vartheta_{12}+r_2)
+y_3y_1(\vartheta_{23}+s_3)(\vartheta_{23}+r_3)
+y_1y_2(\vartheta_{31}+s_1)(\vartheta_{31}+r_1)\\
+y_1y_2y_3(\Delta_3+v_1\vartheta_1+v_2\vartheta_2+v_3\vartheta_3+M_1),
\end{array}
\end{equation}
where $(r_1,r_2,r_3)=(r'_3-1,r'_1-1,r'_2-1)$.
Thus we arrived at the eq.(\ref{eq:main-de}) for $N=3$ case.

\section{Summary and discussions}

In this note, we formulated an explicit differential equation ({\ref{eq:main-de})
 which is expected to determine the instanton partition function $Z_{\rm inst}$
in the presence of the full surface operator in ${\cal N}=2$, $SU(N)$ gauge theory
with $N_f=2N$.
The differential equation is derived as a quantization of the FST equation
of isomonodromy type.

In \cite{Alday:2010vg}\cite{Kozcaz:2010yp}, it was claimed that
the partition function $Z_{\rm inst}$ is the conformal block of the affine
Lie algebra ${SL}_N$ (with the insertion of the $K$-operators).
It is known \cite{Res}\cite{Har} that the KZ equation satisfied by the conformal blocks 
can be interpreted as quantization of a typical isomonodromy system, the Schlesinger equation.
Hence, it is quite natural to expect a direct relation between the formulation of \cite{Alday:2010vg}\cite{Kozcaz:2010yp}
 and the isomonodromy approach here. For instance, the specialization of the primary fields $V_{\chi}$,
$\chi=\kappa \Lambda_1, \kappa \Lambda_{N-1}$ in \cite{Kozcaz:2010yp} for the
simple punctures
agrees with the choice of the spectral type $(N-1,1)$. More precise relations between
these two formulations, in particular the understanding of the mysterious $K$-operators,
will be an important future problem.

Though we have considered the isomonodromy deformation of an operator of the form
(\ref{eq:conn}), it can also be formulated by a scalar differential operator
\begin{equation}
L=\partial_z^N+u_2 \partial_z^{N-2}+\cdots+u_N.
\end{equation}
Then the relation to $W_N$-algebras is also naturally expected (see \cite{KMST}\cite{Wyll}
and references therein).

For the present, our understanding of the relation between $4d$ gauge theory and isomonodromy equation is 
still extrinsic.
In \cite{AGGTV} it was noted  that the linear action of the loop operators (monodromy of surface operators)
on the chiral partition function is independent of the gauge coupling. This observation may be a key ingredient
for more conceptual understanding of the isomonodromic nature of gauge theories and the AGT relation.

\section*{Acknowledgments}
We would like to thank H.~Awata, H.~Kanno, H.~Nagoya, T.~Suzuki, T.~Tsuda and S.~Yanagida for 
valuable discussions. 
The work is supported in part by JSPS grants-in-aid No.21340036 and No.S-19104002.

\section*{Appendix A : Proof of eq.(\ref{eq:Fsol})}
\renewcommand{\theequation}{A.\arabic{equation}}\setcounter{equation}{0}
\renewcommand{\thesubsection}{A.\arabic{subsection}}\setcounter{subsection}{0}

From the equation (\ref{eq:N3prime}),
the coefficients $c_{i,j,0}$ in (\ref{eq:psi-c}) are determined by
\begin{equation}
\begin{array}l
c_{i\! - \!1,j,0} (i\! - \!1\! - \!j\! + \!r_{1}) (i\! - \!1\! - \!s_{1})\! + \!c_{i\! - \!1,j\! - \!1,0} (j\! - \!1\! + \!r_{2}) (i\! - \!1\! - \!s_{1})\\
\! - \!c_{i,j\! - \!1,0} (j\! - \!1\! + \!r_{2}) (1\! + \!i\! - \!j\! + \!s_{2})\! - \!c_{i,j,0} \left(i^2\! - \!i j\! + \!j^2\! + \!i u_{1}\! + \!j u_{2}\right)=0.
\end{array}
\end{equation}
Plugging (\ref{eq:Fsol}) into this equation and rewriting the parameters as
$a_0=u_1-j,\ 
a_1=-i,\ 
a_2=-j,\ 
a_3=-u_2-j,\ 
a_4=u_1-r'_1+1,\ 
a_5=u_1-s_2$, the relation we should prove reduces to
\begin{equation}\label{eq:Frel}
\begin{array}l
(a_{0} a_{1}\! -\!a_{1}^2\! -\!a_{2} a_{3}) F\! -\!(a_{0}\! -\!a_{1}) a_{1} F^{a_1}
\!+\!\frac{(1+a_{0}) a_{2} a_{3} (1+a_{0}-a_{1}-a_{4}) (1+a_{0}-a_{1}-a_{5}) }{(1+a_{0}-a_{1}) (1+a_{0}-a_{4}) (1+a_{0}-a_{5})}F^{a_0 a_2}\\
+\frac{(1+a_{0}) a_{1} a_{2} a_{3} }{(1+a_{0}-a_{4}) (1+a_{0}-a_{5})}F^{a_0 a_1 a_2 a_3}=0,
\end{array}
\end{equation}
where $F=F(a_{0};a_{1},a_{2},a_{3},a_{4},a_{5})$ and $F^{a_0 a_2}=F|_{a_0\rightarrow a_0+1, a_2 \rightarrow a_2+1}$ etc.
Expanding the series $F$ in each term, (\ref{eq:Frel}) can be written as
\begin{equation}\label{eq:ps-om}
\sum_{k=0}^{\infty}\varphi_k\omega_k=0, 
\end{equation}
where $\varphi_k=\prod_{i=0}^k\frac{(a_i)_k}{(1+a_{0}-a_{i})_k}$ and
\begin{equation}
\begin{array}l
\omega_k=
(a_{0} a_{1}\! -\!a_{1}^2\! -\!a_{2} a_{3}) (a_{0}\! +\!2 k)
\! -\!(a_{0}\! -\!a_{1}\! +\!k) (a_{1}\! +\!k) (a_{0}\! +\!2 k)\\
+\frac{(1+a_{0}-a_{1}-a_{4}) (1+a_{0}-a_{1}-a_{5}) (a_{0}+k) (a_{2}+k) (a_{3}+k) (1+a_{0}+2 k)}{(1+a_{0}-a_{1}+k) (1+a_{0}-a_{4}+k) (1+a_{0}-a_{5}+k)}\\
+\frac{(a_{0}+k) (a_{1}+k) (a_{2}+k) (a_{3}+k) (1+a_{0}+2 k)}{(1+a_{0}-a_{4}+k) (1+a_{0}-a_{5}+k)}.
\end{array}
\end{equation}
Then the equation (\ref{eq:ps-om}) follows from an identity
\begin{equation}
\varphi_k \omega_k=\varphi_{k+1} u_{k+1}-\varphi_k u_k, \quad
u_k=k(k+a_0-a_2)(k+a_0-a_3),
\end{equation}
since the infinite sum is terminating: $\varphi_k=0$ for $k> \min(i,j)$. $\square$

\end{document}